\input{aipcheck.tex}
\def\selectedoptions{final}

\documentclass[
   \selectedoptions
  ]{aipproc}

\layoutstyle{6x9}
\SetInternalRegister\hbadness{8000} 

\newcommand\doingARLO[2][]{%
  \ifx\mmref\undefined #1\else #2\fi
}
  
\begin{document}

\title 
      [Meson Spectroscopy in Photo-production at CLAS]
      {Meson Spectroscopy in Photo-production at CLAS}


\author{M. Nozar -- CLAS Collaboration }{
  address={TJNAF, 12000 Jefferson Ave, MS 12H, Newport News, VA 23606},
 }


\copyrightyear {2003}

\begin{abstract}
Photo-production of excited mesons in the $1-2$ GeV mass range decaying via multi-pion or multi-kaon emission has been investigated at the TJNAF\thanks{ This work was supported by the U.S. Department of Energy and The U.S. National Science Foundation.} experiment E01-017 (g6c) in the $4.8-5.4$ GeV photon beam energy range.  The main objective of the experiment is to extract resonance parameters of the produced states by way of a Partial Wave Analysis (PWA) technique.  In this paper, we will focus on the general characteristics of the data distributions in both the neutral and charged 3-pion final states, i.e. $\pi^{+}\pi^{-}\pi^{0}$ and $\pi^{+}\pi^{+}\pi^{-}$.
\end{abstract}
\date{\today}
\maketitle
\vspace*{-7mm}
\section{Introduction}
The observation of an isovector exotic state with $J^{pc} = 1^{-+}$ quantum numbers at $1.6$ GeV by the Brookhaven E852 experiment~\cite{E852_3pi} provided the main motivation to search for such a state using a photon beam at JLab.  This state is one of the signature hadronic states outside the constituent quark model, where either the excitation of the gluons within the hadron directly contributes to its degrees of freedom, giving a state configuration of $(q\bar{q}g)$, or that the state is composed of more than one $q\bar{q}$ pair, i.e. $q\bar{q}q\bar{q}$.  Such a state, if produced in a pion beam via $\rho$ exchange, may also be produced in a photon beam, with the role of the beam and the exchange particle reversed and assuming Vector Meson Dominance (VMD)~\cite{AA_AS_00,AA_PP_98,FC_PP_95}.  Photon beams, as probes for exotic meson production, have not been fully explored so far; the existing data on multi-particle final states are very sparse.  The only relevant data comes from a SLAC bubble chamber experiment which employed a backscattered laser photon beam of $19.5$ GeV average energy~\cite{Condo_91}.  In the low mass region, the data is dominated by $a_2(1320)$ production with no clear evidence for $a_1(1260)$.  In the high mass region, the group claimed evidence for a narrow state at $1.775$ GeV, with possible $J^{pc} = 1^{-}, 2^{-}$, or $3^{+}$ quantum numbers.  Because of insufficient statistics, no PWA was performed on the data.  
\vspace*{-2mm}
\section{Experimental Setup and event selection.}
The data for this analysis was collected during Aug.--Sep. of 2001.  The $40$ nA primary beam of $5.7$ GeV electrons at $100\%$ duty factor and was provided by CEBAF (Continuous Electron Beam Accelerator Facility).  The secondary beam of photons was generated in Hall B via bremsstrahlung radiation, using a $3*10^{-4}$ $r.l.$ radiator and a tagging system identifying photons in the $[20\%-95\%]$ range of the incident electron beam energy~\cite{TAGGER_NIM_99}.  The high flux of $5x10^{6}/\rm sec$ photons (in the top $15\%$ of the photon beam energy) was incident on an $18$ cm long $LH_2$ target.  The Hall B experimental hall houses the CLAS (CEBAF Large Acceptance Spectromenter) detector.  CLAS covers a large solid angle with polar angle coverge in the range $8^{\circ} \leq\theta\leq 145^{\circ}$, and azimuthal angle coverage of $80\%$.  The detector, composed of six independent sectors, provides a toroidal magnetic field, where positively charged particles bend outward and negatively charged tracks bend inward.  Three sets of drift chambers in the radial direction are embedded in the space between the coils, and provide charged particle detection and track reconstruction.  A set of time of flight scintillators are used for charged particle identification, and a set of electromagnetic calorimeters are used for neutral particle detection.  Details of the CLAS detector design and performance are described elsewhere~\cite{CLAS_NIM_00}.

The lack of acceptance in the forward region ($\theta_{lab} < 8^{\circ}$) presents a drawback to any meson spectroscopy program at CLAS in its current configuration.  In addition, with the maximum available photon beam energy of $5.4$ GeV, there is a significant contribution from $t$-channel baryon resonance production.

Both neutral and charged $3\pi$ final states were subjected to similar topological and geometrical cuts. The photon beam energy was selected to be higher than $4.8$ GeV; vertex position cuts were applied to ensure the events originated within the target volume, and vertex timing constraints were imposed to reduce the accidentals between the CLAS and the tagging system.
\vspace*{-2mm}
\section{Data Distributions: $\gamma \lowercase{p} \rightarrow \pi^{+} \pi^{+} \pi^{-} (\lowercase{n})$}
The $\pi^{+}$, $\pi^{+}$, and $\pi^{-}$ were detected in CLAS, and the neutron was reconstructed by way of missing mass.  In order to suppress the $t$-channel baryon background, we have employed two kinematical cuts, a low four-momentum transfer to the $\pi^{+} \pi^{+} \pi^{-}$ cut defined as $-t' \leq 0.4$ GeV$^{2}$, and the other, a small lab polar angle cut imposed on the two positively charged pions defined as $\theta_{lab} \leq 30^{\circ}$.  These two cuts, referred to as ``baryon rejection cuts'', discard mostly events associated with background processes.  In all the plots shown for this final state, the shaded histograms represent the events that passed these two cuts collectively.  

The left plot in Fig.~\ref{fig:t,mm,3pi} shows the missing mass off the $\pi^{+} \pi^{+} \pi^{-}$ for low $-t'$ events.  The neutron peak sits on top of a negligible linearly increasing background, with a signal to background ratio of approximately $9/1$.  A Gaussian plus a $1^{\rm st}$ order polynomial fit to the peak gives a $\sigma$ of $25$ MeV.  The area between the lines around the neutron peak represents the neutron selection cut.  The $-t'$ distribution shown in the middle plot of Fig.~\ref{fig:t,mm,3pi} is fit to an exponential function of the form $f(t') = a\,e^{-b|t'|}$.  The shape of the distribution is consistent with the characteristics of peripheral production, and the exponential constant of $4.4$ $\rm GeV ^{-2}$ is consistent with $\pi$ and $\rho$ exchange~\cite{MG_JM_MV_97}.  In the $3\pi$ invariant mass spectrum shown in the right plot of Fig.~\ref{fig:t,mm,3pi} two enhancements are evident, one in the $1300$ MeV region, and another in the $1600-1700$ MeV range.

\begin{figure}[hbt!]
\begin{minipage}[t]{0.33\linewidth}
\resizebox{9pc}{!}{\includegraphics{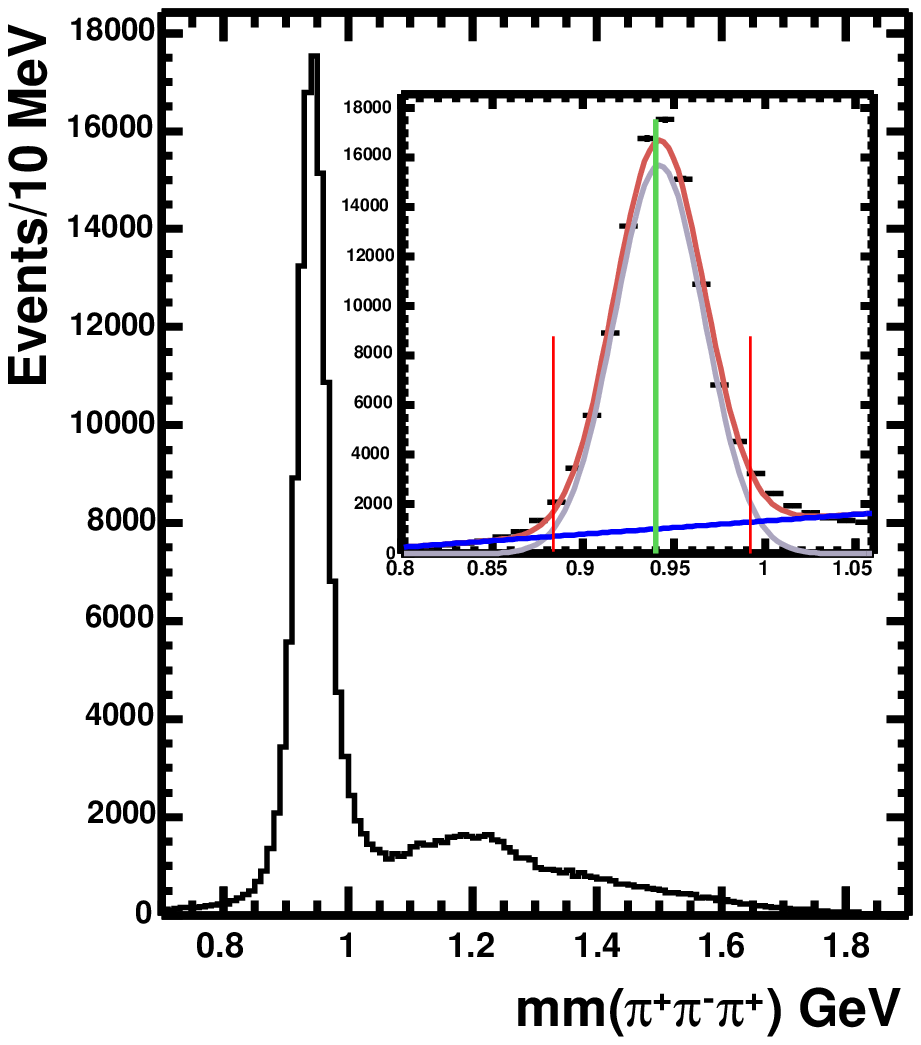}}
\end{minipage}%
\begin{minipage}[t]{0.33\linewidth}
\resizebox{10pc}{!}{\includegraphics{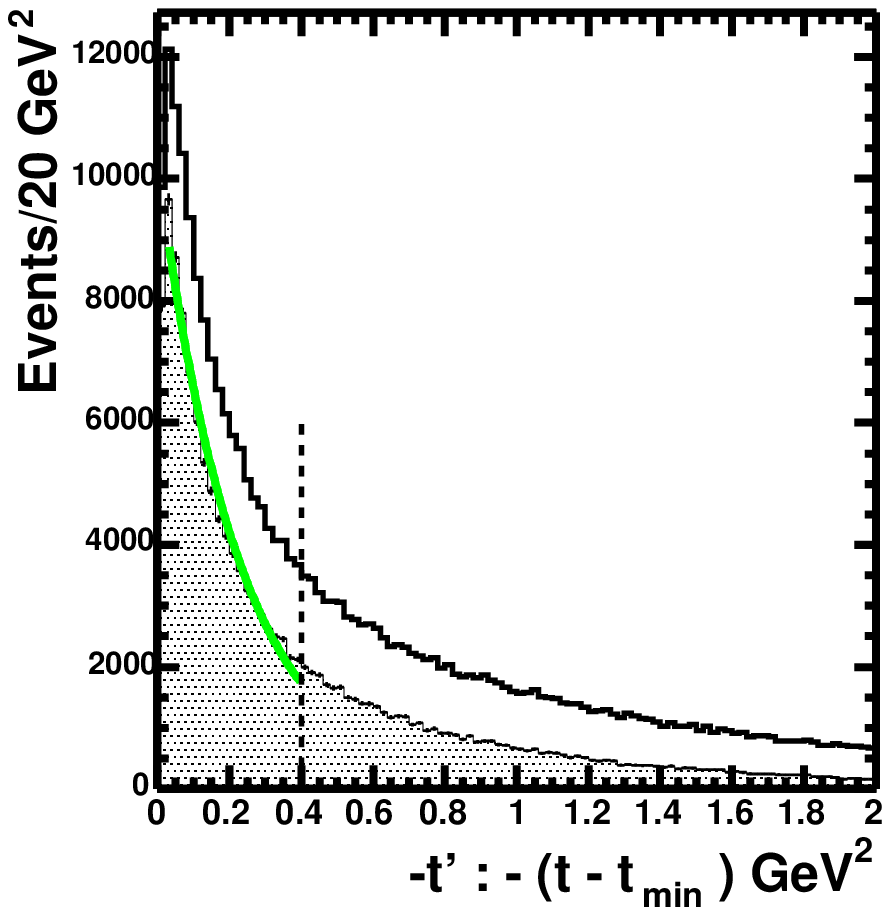}}
\end{minipage}%
\begin{minipage}[t]{0.33\linewidth}
\resizebox{10pc}{!}{\includegraphics{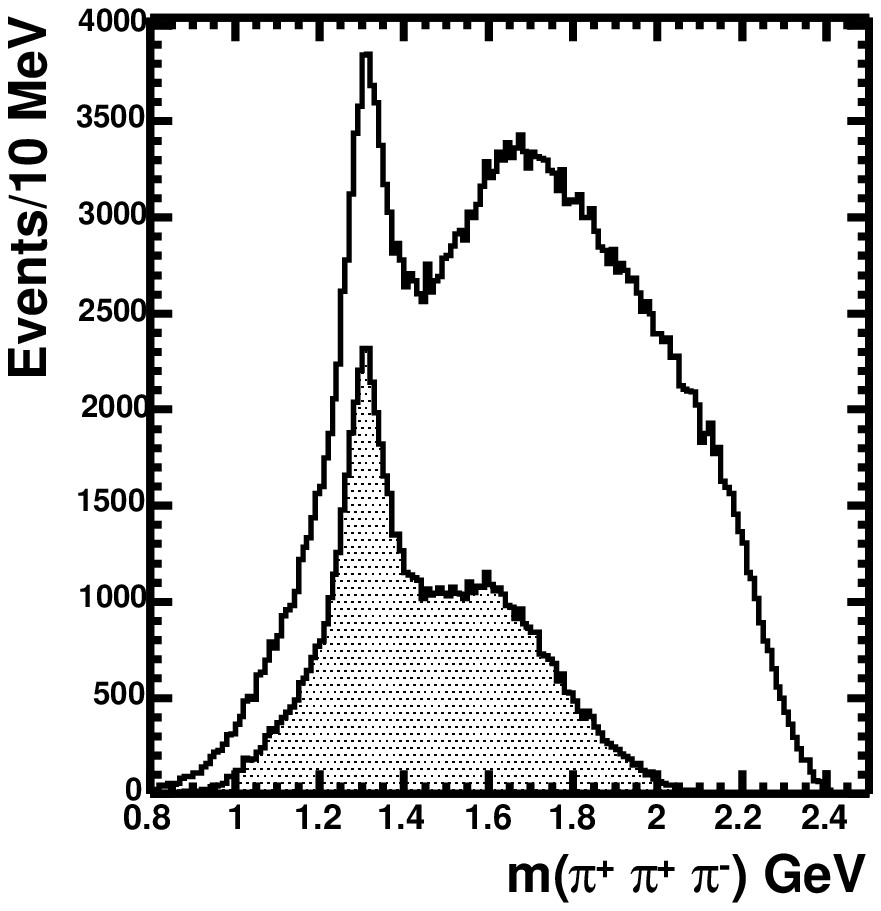}}
\caption{Left: Missing mass off of the $\pi^{+} \pi^{+} \pi^{-}$ for low $-t'$ events. Middle: four-momentum transfer to $3\pi$ for $\pi^{+} \pi^{+} \pi^{-} (n)$ events. Right: $\pi^{+} \pi^{+} \pi^{-}$ invariant mass distribution.}
\label{fig:t,mm,3pi}
\end{minipage}
\end{figure}

\begin{figure}[hbt!]
\includegraphics[width=6in,totalheight=3.8in]{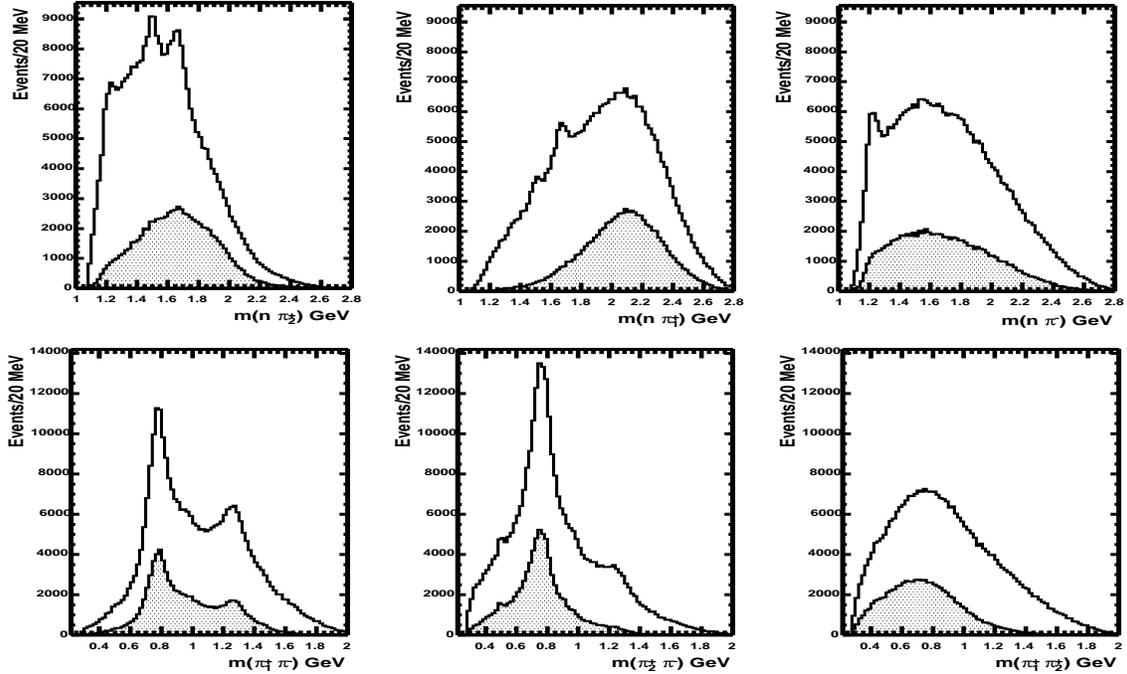}
\label{fig:npi_pipi}
\caption{$n \pi$ (top row) and $\pi\pi$ (bottom row) invariant mass distributions for $\pi^{+} \pi^{+} \pi^{-} (n)$ events.}
\end{figure}

Figure~\ref{fig:npi_pipi} shows all three possible combinations of the $n \pi$ and $\pi\pi$ invariant mass distributions.  In this analysis, the two positively charged pions were sorted based on momentum, with the $\pi^{+}_1$ being the one with higher momentum.  The two $n \pi^{+}$ combinations show peaks around the known baryon resonances, $\Delta(1232)$, $N^{\star}(1520)$, and $N^{\star}(1680)$, while the $n\pi^{-}$ shows a peak around the $\Delta(1232)$ only as is expected due to isospin considerations.  It is clear from the shaded distributions that the baryon resonance peaks are greatly suppressed after the ``baryon rejection'' cuts.  The neutral $2\pi$ effective mass distributions show signals around the mass of the $\rho(770)$ and the $f_{2}(1270)$, as well as a shoulder at the mass of the $f_{0}(980)$.  The doubly-charged $2\pi$ combination doesn't show any distinct peak, indicative of the lack of an isospin $I=2$ state.  
\vspace*{-2mm}
\section{Data Distributions: $\gamma \lowercase{p} \rightarrow \lowercase{p} \pi^{+} \pi^{-} (\pi^{0})$}
The $p$, $\pi^{+}$, and $\pi^{-}$ were detected in CLAS while the $\pi^{0}$ was reconstructed by way of missing mass.  In order to enhance the $3\pi$ mesonic events, a combination of low $-t'$ and $p\pi$ mass cut was employed. The cut was defined as $0.1 \leq -t' \leq 1.0$ GeV$^{2}$, with the $p\pi$ effective masses for all three combination above $1.35$ GeV.  We'll refer to these two cuts, as ``baryon rejection cuts''.  In all the plots shown for this final state, the shaded histograms represent the events that passed these two cuts.

The left plot in Fig.~\ref{fig:Ji_t,mm,3pi} shows the missing mass squared off the $p \pi^{+} \pi^{-}$.  A Gaussian plus a $2^{\rm nd}$ order polynomial fit to the peak around the $\pi^{0}$ gives a $\sigma$ of $41$ MeV for the missing mass resolution.  The region between the lines represents the $\pi^{0}$ selection cut; these events were subjected to kinematical fitting in a later stage.  The $-t'$ distribution is shown in the middle plot of Fig.~\ref{fig:Ji_t,mm,3pi}.  The shaded distribution, after the $\Delta(1232)$ rejection cut is fit to an exponential function, $f(t') = a\,e^{-b|t'|}$, with the resulting $b=1.3$ $\rm GeV ^{-2}$.  The $3\pi$ invariant mass distribution is shown in the right plot of Fig.~\ref{fig:Ji_t,mm,3pi}, with the shaded histogram showing the events which passed the ``baryon rejection cuts''.  In the remaining events, one is able to see enhancement of events at $1300$ MeV and in the $1600-1700$ MeV region.

\begin{figure}[hbt!]
\begin{minipage}[t]{0.33\linewidth}
\resizebox{10pc}{!}{\includegraphics{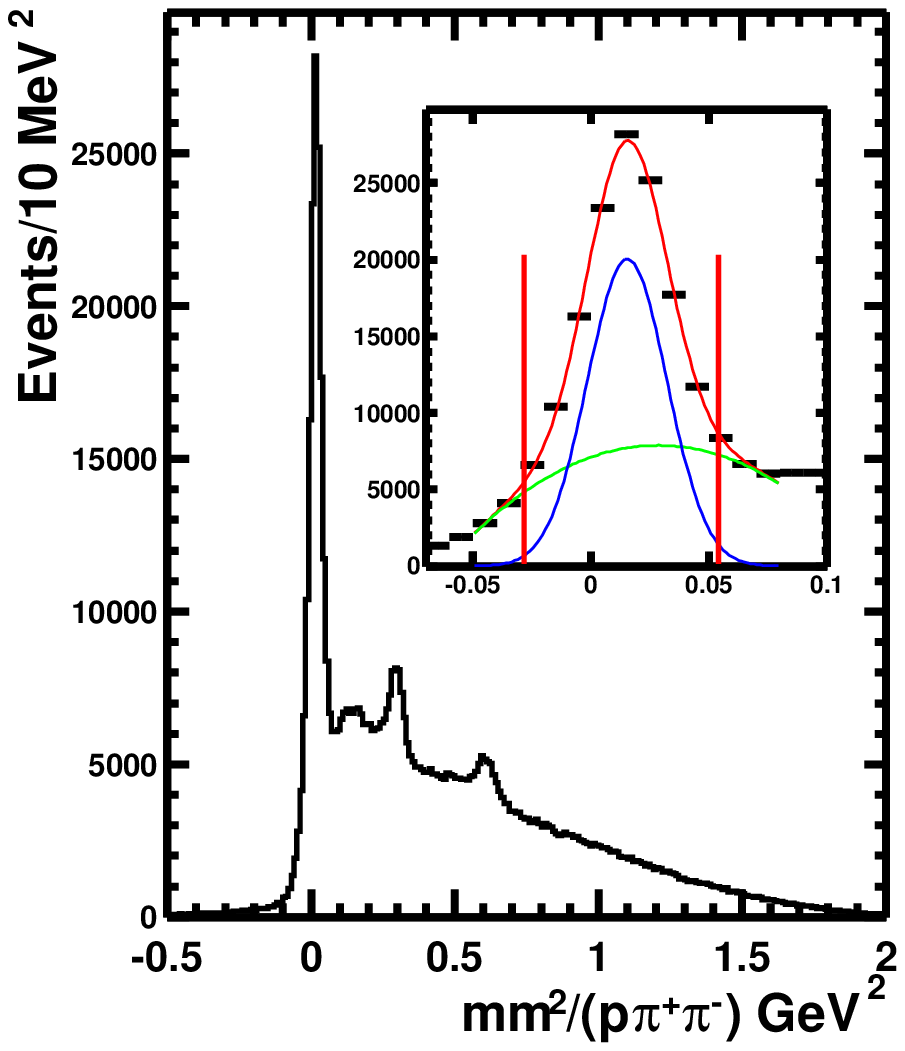}}
\end{minipage}%
\begin{minipage}[t]{0.33\linewidth}
\resizebox{11pc}{!}{\includegraphics{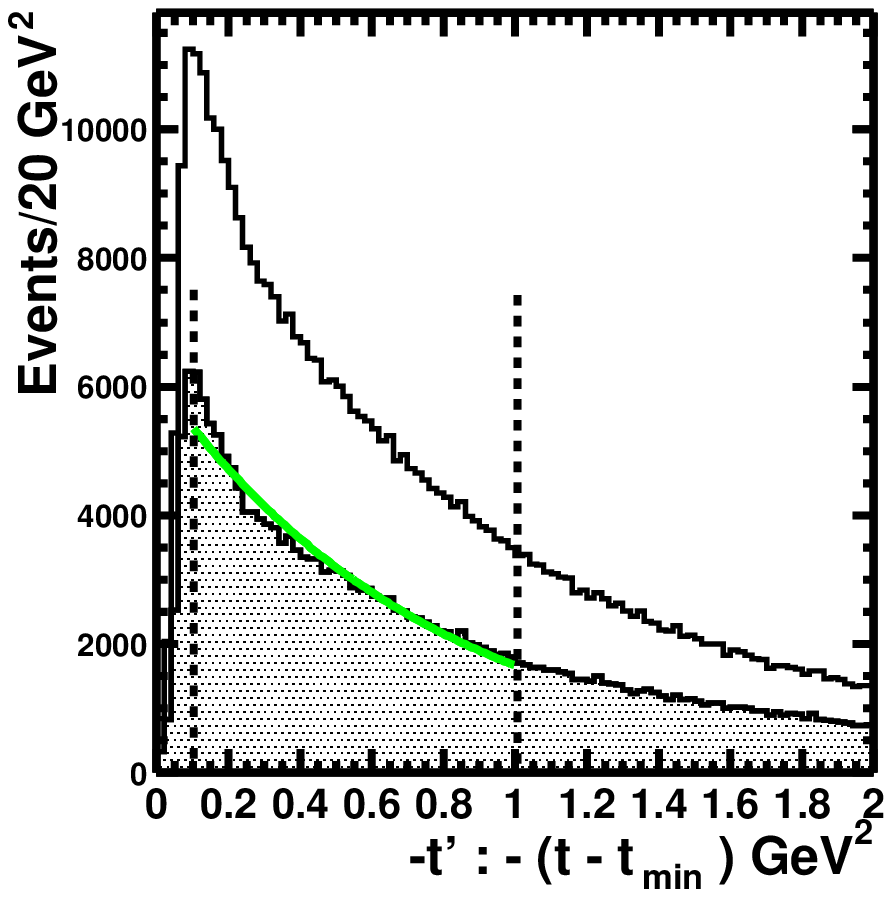}}
\end{minipage}%
\begin{minipage}[t]{0.33\linewidth}
\resizebox{11pc}{!}{\includegraphics{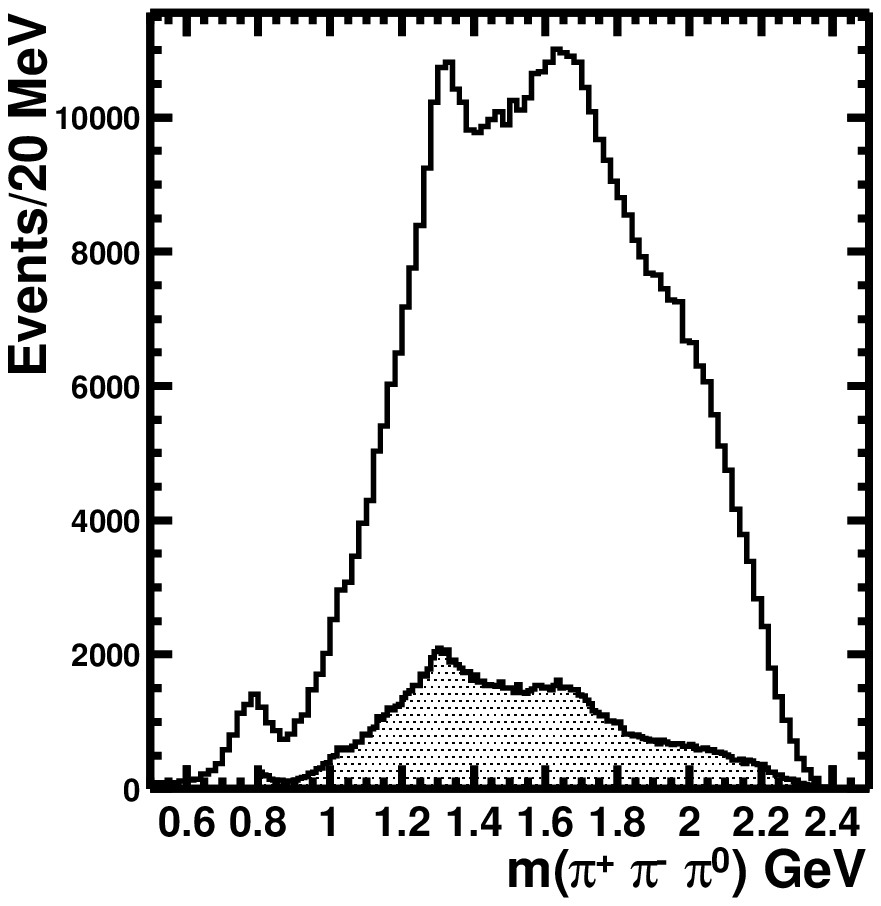}}
\caption{Left: Missing mass off of the $p \pi^{+} \pi^{-}$ for low $-t'$ events. Middle: four-momentum transfer to $3\pi$ for $p \pi^{+} \pi^{-} \pi^{-} (\pi^{0})$ events. Right: $\pi^{+} \pi^{-} (\pi^{0})$ invariant mass distribution.}
\label{fig:Ji_t,mm,3pi}
\end{minipage}
\end{figure}

\begin{figure}[htb!]
\includegraphics[width=6in,totalheight=3.8in]{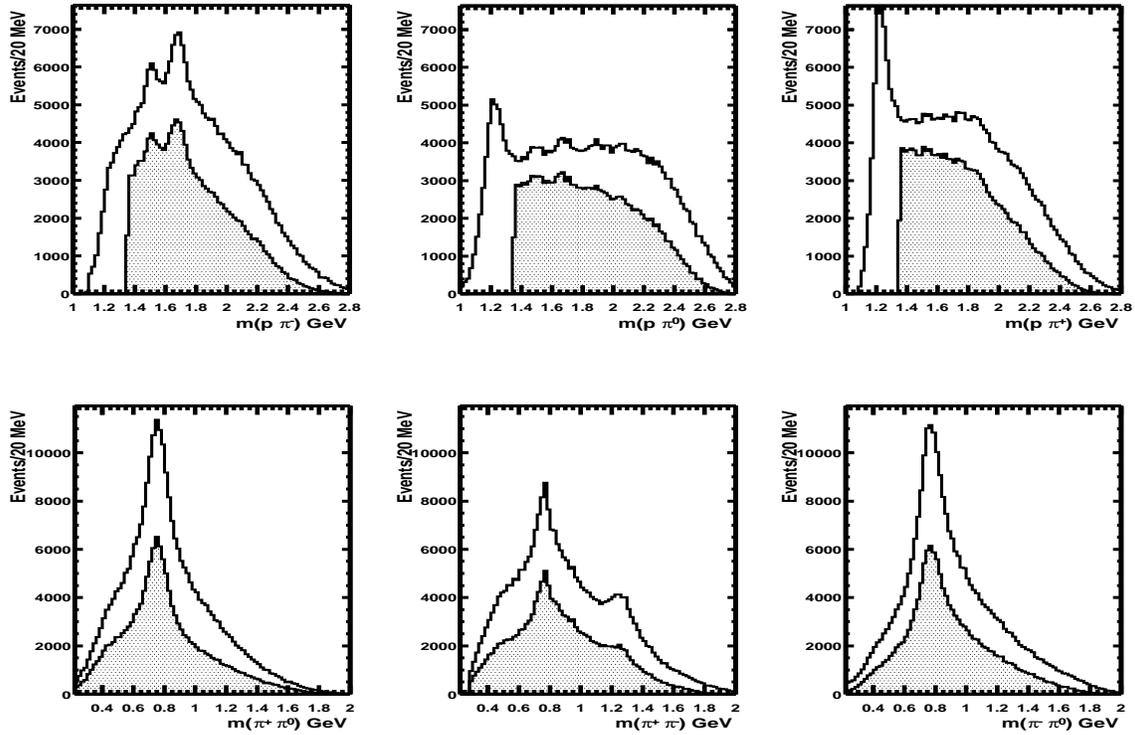}
\label{fig:Ji_ppi_pipi}
\caption{$p \pi$ (top row) and $\pi\pi$ (bottom row) invariant mass distributions for $p \pi^{+} \pi^{-} (\pi^{0})$ events.}
\end{figure}

Figure~\ref{fig:Ji_ppi_pipi} shows all three possible combinations of the $p \pi$ and $\pi\pi$ invariant mass distributions.  Before the ``baryon rejection cuts'', the most prominent feature observed is the peak around the $\Delta(1232)$ mass region.  After the cuts, the peaks around the $N^{\star}(1520)$ and $N^{\star}(1680)$ still remain.  The invariant mass distributions for the $2\pi$ charged combinations, $\pi^{+}\pi^{0}$ and $\pi^{-}\pi^{0}$, show strong peaks at the $\rho(770)$ mass, while the neutral $2\pi$ invariant mass distribution shows enhancements around the $\rho(770)$ and $f_{2}(1270)$ masses.

\section{Summary and Outlook}
We have shown the data quality and the general characteristics of the distributions for both the neutral and the charged $3\pi$ final states from the data collected during the g6c running of CLAS.  The statistics for both sets of events is by a factor of few hundred higher than the existing data.  Partial Wave Analysis on both systems are underway.  
\vspace*{-2mm}
\begin{theacknowledgments}
The invaluable efforts of the Jlab Accelerator staff, the Physics Division, and the g6c group are greatly appreciated.  I am especially grateful to Ji Li for his analysis of the neutral $3\pi$ final state and for providing the figures for this final state.
\end{theacknowledgments}
\vspace*{-2mm}
\bibliographystyle{aipproc}
\bibliography{CIPANP_paper}
\end{document}